\documentclass[preprint]{epl}
\usepackage{amsmath,amssymb} 

\def\araa{ARA\&A}

\def\aaps{A\&AS}

\def\pre{Phys. Rev. E}
\begin{document}

\title{The large-scale organization of chemical reaction networks\\  
in astrophysics}

\shorttitle{Chemical reaction networks in astrophysics}

\author{R. V. Sol\'{e}\inst{1,2,3}\thanks{E-mail \email:ricard.sole@upf.edu}
\and A. Munteanu\inst{1}\thanks{E-mail \email:andreea.munteanu@upf.edu}}

\shortauthor{Sol\'{e} \& Munteanu}

\institute{
\inst{1} ICREA-Complex Systems Lab, Universitat Pompeu Fabra (GRIB), 
Dr. Aiguader 80, Barcelona 08003, SPAIN \\
\inst{2} Santa Fe Institute, 1399 Hyde Park Road, New Mexico 87501, USA.\\
\inst{3} Center of Astrobiology (NASA-associate), Ajalvir Km 4, 
Torrejon de Ardoz, Madrid.
}
                         

\maketitle

\begin{abstract}
The  large-scale organization  of complex  networks, both  natural and
artificial, has  shown the existence of  highly heterogeneous patterns
of  organization. Such  patterns typically  involve  scale-free degree
distributions and small world,  modular architectures.  One example is
provided  by  chemical  reaction   networks,  such  as  the  metabolic
pathways. The  chemical reactions of the Earth's  atmosphere have also
been  shown to give  rise to  a scale-free  network.  Here  we present
novel data analysis on the structure of several astrophysical networks
including  the   chemistry  of  the  planetary   atmospheres  and  the
interstellar medium. Our work reveals that Earth's atmosphere displays
a  hierarchical organization, close  to the  one observed  in cellular
webs.   Instead, the  other astrophysical  reaction networks  reveal a
much  simpler  pattern  consistent  with an  equilibrium  state.   The
implications for large-scale regulation  of the planetary dynamics are
outlined.

\end{abstract}


\section{Introduction}

 The interstellar medium (ISM)  -- gas and micron-sized dust particles
between the stars  -- is the raw material for  the formation of future
generations of stars which may develop planetary systems like our own.
Astronomical  observations of  interstellar and  circumstellar regions
have lead  to the  identification of well  over one  hundred different
molecules, most of them being organic in nature \cite{EC00}. Motivated
by  these   discoveries,  astrochemistry  --  the   chemistry  of  the
interstellar  gas --  has developed  into an  active research  area of
astrophysics and detailed chemical models can now be constructed which
reconstruct the history and role  of the ISM in the evolutionary cycle
of the galaxy \cite{herbst01}.   Crucial to modeling chemical kinetics
in the interstellar medium, the UMIST kinetic database \cite{LTMM2000}
consists in the chemical reactions relevant to astrochemistry.

In  view of the  increasing data  on the  chemical composition  of the
solar  system's  planets  from  latest planetary  missions,  there  is
growing  interest  in the  astrophysical  community  for modeling  the
weather and  atmospheric chemistry  of the neighboring  planets.  Such
modeling   provided  extensive   chemical   reaction  networks   (CRN)
\cite{YDM99} that  expect confirmation from  future planetary missions
and further modeling.  Using  the general approach of complex networks
\cite{DM03} we  explore here the large-scale topology  of the chemical
networks   associated  to  the   interstellar  medium   and  planetary
atmospheres. As will be shown,  two basic types of networks are found,
being associated with the presence or absence of life.


\begin{figure}[t]
\begin{center}
\includegraphics[scale=0.5]{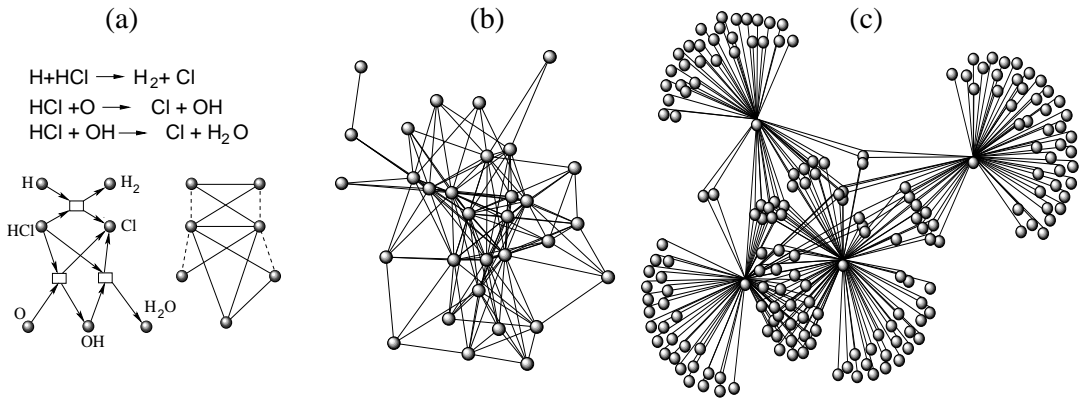}
\caption{{\sl (a)} Defining the reaction graph for a set of three 
reactions indicated  in the upper part: directed  bipartite graph {\sl
(l.h.s)  } and  undirected substrate  graph {\sl  (r.h.s)},  where the
links  between reactants and  between products  {\sl (dashed)}  may be
considered or  disregarded. {\sl (b)}  The overall reaction  graph for
Martian atmosphere is shown.  {\sl  (c)} A subgraph of Earth's CRN. See
text for details.}
\label{EPLREACTIONSPK}
\end{center}
\end{figure}

\section{Reaction graphs}

A CRN  can be viewed as a  graph where chemical species  are nodes and
edges  represent conversion  between  chemicals. The  most simple  and
typical   representations   of  a   reaction   graph   are  shown   in
fig.~\ref{EPLREACTIONSPK}a as  a directed bipartite  graph (l.h.s) and
an undirected substrate graph (r.h.s). We have chosen to disregard the
connections  between  the reactants,  on  one  side,  and between  the
products,    on    the   other    side    (the    dashed   lines    in
fig.~\ref{EPLREACTIONSPK}a).   In  fig.~\ref{EPLREACTIONSPK}, we  also
show  as examples the  undirected substrate  graphs associated  to the
chemical reactions  of the  Martian atmosphere ({\sl  panel b})  and a
subgraph of the Earth network ({\sl panel c}).

 Two basic features common to many complex networks, from the Internet
\cite{barabasialbert,caldarelli}        to        cellular        nets
\cite{jeongprot,jeongmet}   are   their   {\em  scale-free}   topology
\cite{albertbarabasi}         and         small-world        structure
\cite{wattsstrogatz,newmanSW}. The first states that the proportion of
nodes  $P(k)$ having  degree  $k$ decays  as  a power  law $P(k)  \sim
k^{-\gamma}\phi(k/\xi)$,  with $\gamma \approx  2-3$ for  most complex
networks  \cite{albertbarabasi,barabasialbert,amaral}   and  with  the
function  $\phi(k/\xi)$ introducing a  cut-off at  some characteristic
scale $\xi$.  The power-law distribution has no natural scale and from
here  networks with  such  distributions are  called scale-free.   The
second refers to a web characterized by a very small diameter (average
shortest path between any two  vertices) along with a large clustering
\cite{wattsstrogatz,newmanSW}. The {\sl  average path length} $\langle
L \rangle$ is  the average minimum distance $d(i,j)$  between any pair
$(i,j)$ of vertices:
\begin{equation}
\langle L \rangle = \frac{1}{V(V-1)} \sum_{\forall i,j} d(i,j),
\end{equation}
\noindent
The {\sl clustering coefficient} $C_{\rm  i}$ associated to a node $i$
characterizes the density of  links in its neighborhood (the fraction
of  its neighbors that  are also  neighbors of  each other).   It is
defined as $C_{\rm i} =  2V_{\rm i}/N_{\rm i}(N_{\rm i}-1)$, being the
ratio  between the  total number  of  links, $V_{\rm  i}$ between  its
nearest neighbors, $N_{\rm  i}$ and the total number  of all possible
edges  between  all  these  nearest neighbors.   The  average  value,
$\langle C \rangle$  is the clustering coefficient of  the network and
may  be   considered  as  an  indicator  of   a  potential  modularity
\cite{ravasz02}, as discussed  below. For a small world,  we have $<L>
\approx <L_{rand}>$ whereas $<C> \ll <C_{rand}>$, where the index {\em
rand}  refers  to  the   random  counterparts  of  the  network  under
consideration \cite{wattsstrogatz,newmanSW}.

\begin{table}[t]

\caption{Network characteristics: V = vertices; E = edges; $\langle K \rangle$ = mean degree; $\langle L \rangle$ = averaged shortest path; $\langle C
\rangle$ =  mean clustering; $r$  = assortativity.  See text  for more
details. ISM:  Interstellar medium; HC :  hydrocarbon chemical network
of the giant planets} \vspace{0.2cm}
\begin{center}
\begin{tabular}{|c|c|c|c|c|c|c|c|c|c|c|}
\hline
\hline

&  V &  E &  $\langle K  \rangle $  & $\langle  L \rangle$  & $\langle
L_{rand} \rangle$ &  $\langle C \rangle$ & $  \langle C_{rand} \rangle
$& r & $r_{rand}$ & Modular\\

\hline
Earth & 248 & 778 & 6.27 & 2.75 & 3.20 & 0.31 & 0.025 & -0.31 & -0.006 & YES \\ 
Mars & 31 & 144 & 9.29 & 1.89 & 1.73 & 0.61 & 0.31 & -0.10 & -0.007 & NO \\
Titan & 71 & 396 & 11.16 & 2.08 & 1.98 & 0.55 & 0.16 & -0.17 & -0.03 & NO \\
Venus & 42 & 175 & 8.33 & 2.07 & 1.94 & 0.59 & 0.20 & -0.14 & -0.06 & NO\\
HC & 39 & 270 & 13.85 & 1.65 & 1.64 & 0.68 & 0.37 &  -0.26& -0.06 & NO\\
\hline
\hline

ISM & 400 & 6102 & 30.51 & 1.99 & 2.01 & 0.52 & 0.07 &
-0.24 & -0.006 & NO \\
\hline \hline
E.coli & 741 & 2310 & 6.24 & 3.02 & 3.82  & 0.183 & 0.008 & -0.17 & 0.004 & YES\\
\hline \hline
\end{tabular}        
\end{center}
\end{table}

\section{Astrophysical networks}

A first analysis of the  CRNs of planetary atmospheres was included in
\cite{gleiss01}   who  considered  the   chemical  data   reported  in
\cite{YDM99}  and concluded  that  the Earth  chemical  network has  a
scale-free  degree distribution. In  the quantum  chemistry framework,
Patra et al. \cite{patra97}  proposed a reaction mechanism for certain
interstellar reactions using a graph-theoretical approach. The present
work is  a thorough exploration from the  complex networks perspective
of the complete data set  concerning the planetary atmospheres and the
astrochemical  network,  including the  hydrocarbon  chemistry of  the
jovian  planets  \cite{YDM99}.   Among  the  planetary  networks,  the
chemical  reactions associated  to the  Titan's atmosphere  were taken
from \cite{toublanc95},  involving 99 more chemical  reactions than in
\cite{YDM99}.

We have  determined the average characteristics of  the planetary CRNs
and of the  astrochemical UMIST network.  These measures  are given in
table~I.  A  first  result   from  this  analysis  (consistently  with
\cite{gleiss01}) is that all networks are small worlds. The sparser of
these  nets  is  Earth,  with  an  average  degree,  path  length  and
clustering similar to those displayed by the metabolic network of {\em
E. coli}, which has a similar size.

Column 9 gives the assortativeness 
\cite{newman02} defined as:
\begin{equation}
r =  \frac{E^{-1}\sum_i j_ik_i  - [ E^{-1}  \sum_i \frac{1}{2}  (j_i +
k_i)]^2}{E^{-1} \sum_i  \frac{1}{2} (j_i^2  + k_i^2) -  [E^{-1} \sum_i
\frac{1}{2} (j_i + k_i)]^2},
\label{eq:assort}
\end{equation}

\noindent 
where $j_i$,$k_i$  refer to the  degrees of the  nodes at the  ends of
$i^{th}$  link,  with  $E$  being   the  total  number  of  edges  and
$i=\overline{1,E}$.   The  assortativity  coefficient  quantifies  the
propensity of  nodes to connect  to nodes of similar  degree.  Complex
networks  tend  to  be  disassortative  (i.   e.   $r<0$),  reflecting
low-degree  nodes'  tendency to  be  connected  to high-degree  nodes,
consistently       with      the      patterns       displayed      in
fig.~\ref{EPLREACTIONSPK}c.  As  it results from  table~I, the Earth's
atmosphere network and the  astrochemical network present a pronounced
disassortative  character, while  the rest  of the  planetary networks
show more neutral $r$ values.

\begin{figure}
\begin{center}
\includegraphics[scale=0.52]{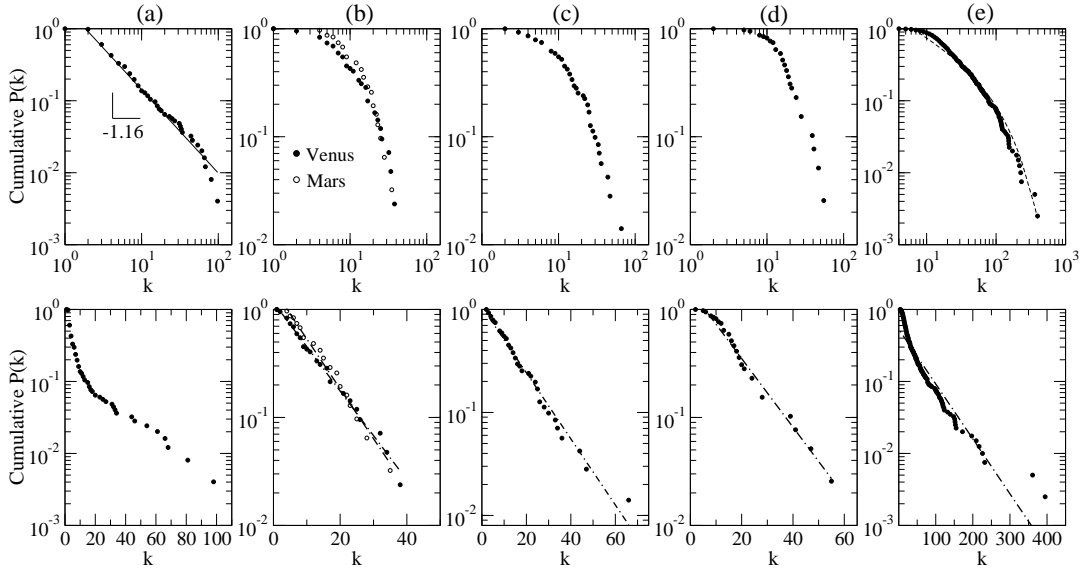}
\caption{
Cumulative   degree  distribution   $P_{cum}(k)$  for   the  planetary
atmospheres  and astrochemical reaction  networks represented  both in
log-log  {\sl  (upper panels)}  and  linear-log  {\sl (lower  panels)}
scales. {\sl  (a)} Earth; {\sl (b)}  Mars and Venus;  {\sl (c)} Titan;
{\sl (c)}  Hydrocarbon chemistry;  {\sl (e)} Interstellar  medium.  It
reveals a scale-free distribution  for the Earth network with exponent
$\gamma  =  2.16$  --   (solid  line  fit);  exponential  distribution
(dashed-dot  lines)  for the  other  networks,  with the  interstellar
medium  network  falling slower  in  the  tail  than expected  for  an
exponential  decay.   This  CRN  is  better fitted  by  a  broad-scale
distribution  $P_{\rm cum}(k)  \sim k^{-\gamma}  \exp(-k/\xi)$ (dashed
line fit). }
\label{PKdists}
\end{center}
\end{figure}

The heterogeneous  character of the degree  distributions is displayed
in  fig.~\ref{PKdists}  using the  cumulative  distribution for  these
networks.  It is  defined as $P_{\rm cum} (k)  = \int_k^{\infty} P(k')
dk'$ which  gives $P_{\rm cum} (k) \sim  k^{-\gamma+1}$ for scale-free
nets.   Both    log-log   and   linear-log   plots    are   shown   in
fig.~\ref{PKdists}, as they are indicators of the type of connectivity
distribution \cite{amaral}.  The Earth network  (fig. 2a) is  the only
one  displaying  a  scale-free  topology.  For the  rest  of  the  the
planetary CRNs, the good exponential fit consisting in a straight line
in  the linear-log  plot reveals  single-scale networks.  From network
considerations alone,  the richness of  the Titan's chemistry  and the
poverty of Mars' are recovered  from the chemical character of the two
CRNs,  the  former  being  reducing  in nature  while  the  second  is
oxidized.  While  a completely  oxidized atmosphere allows  no organic
chemistry and therefore, no complex CRN, a reducing atmosphere such as
that of the outer planets and the interstellar medium implies no limit
for the  complexification of carbon chains in  organic chemistry. From
this perspective, the richness of the interstellar medium chemistry is
reflected  in  table~I:  a   large  highly-connected  network  with  a
broad-scale   distribution  (Fig.~\ref{PKdists}d).   This  observation
discards  a simple  explanation for  Earth's scaling  based on  a high
chemical diversity.

\begin{figure}
\begin{center}
\includegraphics[scale=0.65]{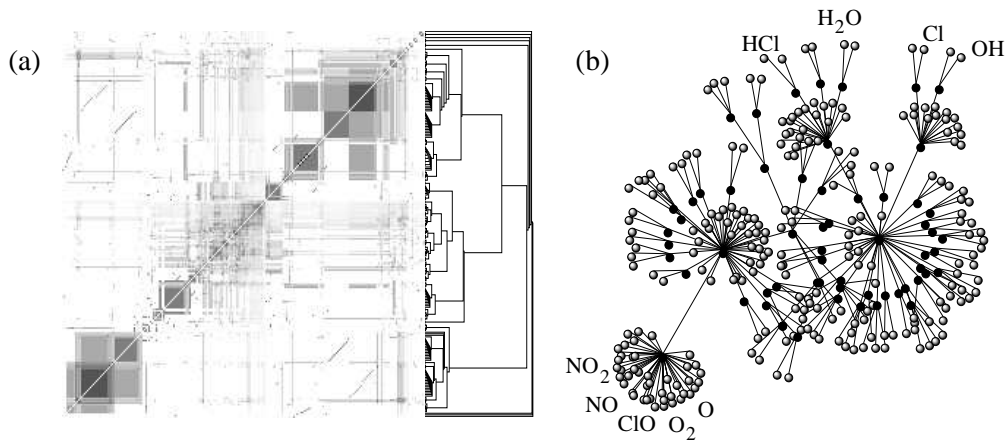}
\caption{Modular  structure   of  Earth's  atmosphere  CRN.   {\sl (a)}  
Topological  overlap
matrix and  the associated dendrogram.   {\sl (b)} The  community tree
transformed into a hierarchical community graph on the basis of the HS
index: white nodes  are the nodes of the original  graph and the black
nodes are the communities to  which they belong. }
\label{earthmoduls}
\end{center}
\end{figure}

Another property that  seems to be common to  many complex networks is
modularity   \cite{guimera03,krause03,holmehuss03}.   Several  methods
provide valuable insight on the existence of such community structures
\cite{newmangirvan}.   One of them  is {\em  hierarchical clustering},
based   on  identifying   similarities  between   nodes   through  the
topological overlap matrix. The topological overlap of a pair of nodes
$(i,j)$  is defined  as $O(i,j)  = J(i,j)/{\rm  min}\{k_{\rm i},k_{\rm
j}\}$, where  $J(i,j)$ denotes the number  of nodes to  which both $i$
and $j$ nodes are connected. The denominator gives the smallest degree
of  the  pair  of  nodes  $(i,j)$. Hierarchical  clustering  uses  the
previous matrix to  identify modules based on the  assumption that the
larger the overlap between two  nodes, the higher the probability that
they belong  to the same  module.  The algorithm repeatedly  finds the
highest topological  overlap, grouping together  the corresponding two
nodes into a new aggregated  node and performing a weighted average of
the   corresponding   row   and   column   in   the   overlap   matrix
\cite{ravasz02},  and  computes  the  new matrix.   The  procedure  is
applied  until the  matrix collapses  into a  single value,  having as
final result an alternative representation of the network as a tree or
dendrogram.   

The result of the application of this method to the Earth's atmosphere
CRN is shown in fig.~\ref{earthmoduls}a, revealing certain communities
of nodes  characterized by higher topological  overlap (dark regions).
In particular, we have noticed that  the main modules -- the two black
squares in  panel (a) --  include predominantly the reactants,  in one
module, and the  products, in the other module,  involved in reactions
with $\rm  OH$ (or $\rm  Cl$) and having  as one of the  products $\rm
H_2O$ (or $\rm HCl$). The  community structure is a consequence of the
action of the  most reactive free radicals of  the Earth's atmosphere,
Cl and OH,  capable of reacting with almost all  other elements due to
their  unpaired electron. Together  with the  rest of  the atmospheric
radicals, such as NO, ${\rm NO_2}$, ${\rm NO_3}$, ClO, they constitute
the  source of  oxidizing  power of  the  atmosphere, determining  the
lifetime and the abundance of  trace species and acting as atmospheric
regulators. Among all these  oxidizing agents, the hydroxyl radical is
by far  the primary atmospheric  oxidant, while ${\rm O_2}$  and ${\rm
O_3}$, in  spite of being  the most abundant oxidants,  are relatively
unreactive.   As  a  network   measure  of  reactiveness,  the  node's
out-degree recovers the reactiveness hierarchy of the oxidizing agents
(decreasing out-degree):  OH, Cl,  ${\rm NO_2}$, O,  ${\rm O_2}$, ${\rm
O_3}$,  NO, ${\rm HO_2}$  and ClO.   The Earth's  oxidizing atmosphere
characterized by a great  chemical disequilibrium, with both oxidizing
and reducing gases in a highly reactive mixture, is the unique imprint
of the  terrestrial atmosphere  compared to the  planets of  the Solar
System,  and  these characteristics  are  recovered  also through  the
analysis of its CRN.

We  have also  analyzed  the measure  of  betweenness centrality  (BC)
\cite{newman01b}  which counts  the fraction  of shortest  paths going
through  a given  edge. The  edge of  highest BC  is likely  to bridge
community structures and thus by calculating and subsequently removing
this edge  through an iterative  algorithm, one obtains  a reasonably
accurate   unfolding   of  the   modular   structure   of  the   graph
\cite{newmangirvan}.  The removal process  transforms the graph into a
binary  tree similar to  the dendrogram  from fig.~\ref{earthmoduls}a,
but formed of bifurcation nodes representing modules and its branches,
connected  to submodules or  individual nodes  of the  original graph.
For the  purpose of extracting the hierarchical  modularization we use
the Horton-Strahler (HS) \cite{guimera03}.  From the bottom to the top
of the tree, the HS index changes  only when it joins a community of a
similar index: the individual nodes  (HS=1) join to form a group (with
HS=2), which in  turn join other groups to form  a second level (HS=3)
and so  on.  Thus, the communities  of equal HS indexes  form the same
modularization  level.   In  fig.~\ref{earthmoduls}b  we  include  the
associated  hierarchical  level-tree   as  it  results  from  grouping
together communities  of equal HS  index. The white  circles represent
the nodes of the CRN graph, while the black ones represent the modules
to which they  belong. The most relevant modules  are clearly revealed
in this figure.  The two fans  in the upper part of the figure include
the products (r.h.s. fan)  and the reactants, respectively (l.h.s fan)
from reactions  with $\rm  OH$ (or $\rm  Cl$), as discussed  above. We
draw the attention  also on the module visible in  the lower left part
of the figure.  It consists mainly of the highly connected core of the
reactants participating in termolecular  (type A + B $\xrightarrow{M}$
AB)  and photochemical  reactions (A  + hv  $\rightarrow$ B  +  C). It
includes also the  majority of the oxidant agents,  some of them being
depicted in fig.~\ref{earthmoduls}b.   The neighboring module includes
mainly  the  products  of significant  degree  resulting  from
reactions.

\section{Discussion}

Within  cellular networks,  metabolic  pathways are  one  of the  most
relevant components of life \cite{NetworkBiologyNature}. Such networks
are  defined  at  the  microscopic, cell-level  scale,  whereas  those
considered  in our  study deal  with vast  spatial  scales.  Moreover,
cellular networks  result from biological evolution,  whereas the CRNs
studied  here are  generated  from mechanisms  that  seem to  strongly
depart from  this scenario, although  natural selection seems to  be a
key    mechanism   for   the    evolution   of    Earth's   atmosphere
\cite{Lenton1998}.   Other  CRN's  have single-scaled  or  broad-scale
structure.

Earth's atmosphere  is a clear  exception to this rule.   Our analysis
shows that, together with a  broad degree distribution, the CRN of our
planet is  also rich in correlations. This  is particularly remarkable
in terms of the  presence of a well-defined hierarchical organization,
as shown  by its modular, nested  architecture. What is  the origin of
such  difference?   One  clear  candidate  is  the  strong,  nonlinear
coupling  between  atmosphere  and   biosphere.   As  pointed  out  in
\cite{lovelock,lovelock03}, the atmosphere is the face of a planet and
it tells  if it  is alive  or dead. Its  chemical composition  and its
departure  from  a  near-equilibrium  state are  consistent  with  the
presence  of life.  Our  analysis gives  strength to  this conjecture,
since the topological organization  of Earth's atmosphere displays the
hierarchical patterns observed in other living structures. In this
context, it is generally accepted \cite{lovelock03} that our planet is
able  to self-regulate  its  climate and  keep  a chemically  unstable
atmosphere constant  and appropriate for life  \footnote{In spite that
our Sun has increased its output  of heat by about $25-30 \%$ over the
last 2500  Myr}. It does  so by using  the incoming energy  (light) to
recycle  the available  chemicals through  both positive  and negative
feedbacks. Well-defined cycles can  be identified and regulation works
over a wide  range of conditions.  These are  all characteristics of a
metabolism \cite{Morowitz}.
 
The dynamics and composition of the  atmosphere of a given planet is a
consequence  of both  dynamical and  historical  constraints. Physical
factors strongly influence the final pattern at the global scale.  But
not less  important seems the role played  by historical contingencies
and  histeretic processes,  which can  irreversibly modify  a planet's
climate.   Earth,  Venus and  Mars  all  had  water soon  after  their
formation 4.5  billion years ago.  Venus experienced a  global runaway
greenhouse  effect  about 3  billion  to  4  billion years  ago.  Mars
followed  a different  path  towards a  runaway  cooling.  The  common
pattern of organization of both  Mars and Venus CRN confirms that lack
of biosphere  leads to a simple,  equilibrium set of  reactions with a
well-defined,  single-scale  topology.   Instead,  the presence  of  a
mechanism injecting reactive components into the reaction pool might 
eventually generate  a complex network not unlike the ones seen
in living  structures at the  small scale. Future work  should explore
the use of kinetic models to test this conjecture.

\begin{acknowledgments}
We thank P. Stadler, H. Jeong, L. A. Barab\'asi and E. Ravasz
for  providing useful  data.  Special  thanks to  Pau  Fern\'andez and
Sergi Valverde for help at different stages of this work and to Franck
Selsis for  helpful discussions.  This  work was supported by  a grant
BFM2001-2154 and by the Santa Fe Institute (RVS).
 
\end{acknowledgments}

\end{document}